\pdfoutput=1
\documentclass[floatfix,reprint,aps,prc,showpacs,showkeys,eqsecnum,amsmath,amsfonts,amssymb,twocolumn,groupedaddress,superscriptaddress,footinbib,noeprint]{revtex4-2}

\usepackage{graphicx}    

\usepackage{float} 

\usepackage{dcolumn}     
\usepackage{bm}          
\usepackage[hyperindex,breaklinks]{hyperref}

\usepackage{amsmath} 
\usepackage{amssymb} 
\usepackage{amsfonts} 
\usepackage{latexsym}
\usepackage{enumitem} 

\usepackage{booktabs} 

\usepackage{cleveref}

\usepackage[english]{babel}
\usepackage[expansion=all,babel]{microtype}

\usepackage[usenames]{xcolor}

\usepackage{mathtools}

\usepackage{multirow}

\usepackage{rotating}

\newcommand{\be}{\begin{equation}}
\newcommand{\ee}{\end{equation}}
\newcommand{\ba}{\begin{eqnarray}}
\newcommand{\ea}{\end{eqnarray}}

\newcommand{\sat}[1]{{#1}_{\mathrm{sat}}}
\newcommand{\slfrac}[2]{\left.#1\middle/#2\right.}

\usepackage{txfonts}

\makeatletter
	\newcommand{\vast}{\bBigg@{2.85}}
\makeatother

\begin{document}

\preprint{Grant\#}
\title{On the Inner Crusts of Neo-Neutron Stars: exotic light nuclei, diffusional and thermodynamical stability}

\author{Mikhail V. Beznogov}
\email{mikhail.beznogov@nipne.ro}
\affiliation{National Institute for Physics and Nuclear Engineering (IFIN-HH), RO-077125 Bucharest, Romania}

\author{Adriana R. Raduta}
\email{araduta@nipne.ro}
\affiliation{National Institute for Physics and Nuclear Engineering (IFIN-HH), RO-077125 Bucharest, Romania}

\date{\today}

\begin{abstract}
	Based on an extended nuclear statistical equilibrium model, we investigate the properties of non-accreted crusts of young and warm neo-neutron stars, i.e., of finite-temperature inhomogeneous dense matter in beta equilibrium.
    An interesting feature is the appearance, in the deep inner crust, of an extensive and almost pure layer of neutron-rich light nuclei that extends up to the density of the transition to homogeneous matter.
    Most probably, this layer emerges due to translational degrees of freedom of the nuclei.
	If confirmed, it will significantly impact the transport and elastic properties of the crust and its crystallization process.
	Then, we demonstrate that our inner crust is stable with respect to the diffusion of ions, which is in contrast with some of the predictions made in the literature for cold crusts.
    Finally, we show that clusterization completely exhausts the density instabilities that affect sub-saturated nuclear matter.
\end{abstract}

\maketitle
\section{Introduction}
\label{sec:Intro}

An important stage of the evolution of neutron stars (NSs) is the neo-NS phase~\cite{BPRR_2020}.
It begins at the end of the proto-NS phase, when the star's temperature drops below $\approx 2$~MeV, and its material becomes transparent to neutrinos, and lasts for about a day.
While NS cores acquire their final composition and size during the proto-NS phase, NS crusts form and crystallize~\cite{BPRR_2020,Fantina_2020,Carreau_2020a} during the neo-NS stage.
The crust crystallization epoch roughly coincides with the time when the nuclear statistical equilibrium (NSE) freezes, which means that subsequent modifications of the crust composition can only occur through a limited subset of nuclear reactions and at a slow pace~\cite{Chamel_2008,PC_2021}.
Consequently, in reality, an NS crust might never reach the ``cold catalyzed'' state, which is usually considered a standard paradigm for mature isolated NSs~\cite{HPY07,Chamel_2008}.
For accreting NSs, the initial composition of the crust is also important, especially in the case of partial accretion~\cite{Suleiman_2022}.
Thus, since the crust formation process sets the initial composition, it is necessary to investigate it in more details.
Such an investigation requires dedicated simulations that combine thermal evolution with evolution of chemical composition and mechanical structure. 
To our knowledge, no such studies are available so far.

The cooling and contraction of the envelope and the outer crust was addressed in Ref.~\cite{BPRR_2020}. As a further step, in this paper we consider the nuclear composition of warm inhomogeneous matter with densities lower than the nuclear saturation density ($\sat{n} \approx 0.16~\mathrm{fm}^{-3} \approx 2 \cdot 10^{14}~\mathrm{g/cm^3}$) and in beta equilibrium.
We also discuss the questions of the thermodynamic stability of NS crusts and the stability with respect to diffusion of ions (buoyancy).
The presented results were obtained employing an extended NSE (eNSE) model~\cite{Raduta_AA_2025} and a code that we developed specifically for this purpose.
In addition, the composition of the crust in the low temperature limit is confronted with published results corresponding to cold catalyzed matter.

\section{Formalism}
\label{sec:NSE}

For densities lower than $\sat{n}$ and temperatures lower than a dozen MeV, stellar matter consists of a mixture of nuclei (clusters, ions), nucleons, leptons, and photons.
Charged leptons are essential to compensate for the positive charge of the protons.
As the thermal excitation of muons is negligible, local charge neutrality implies that $n_p=n_e$, where $n_e$ stands for the (net) number density of electrons.
The only significant interaction between electrons and protons is the electrostatic one.
The electrons and photons interact weakly and are treated as ideal Fermi and Bose gases, respectively.
Neutrinos are not necessarily in equilibrium with the rest of matter and are disregarded here.

In NSE, which is expected to hold for temperatures $T \gtrapprox 0.3-0.4$~MeV~\cite{Wiescher_2018}, the abundance of $(A,Z)$-nuclei (with $A$ being the mass number and $Z$ being the charge number) is determined only by their intrinsic properties, such as internal partition function $z_{A,Z}^{\mathrm{int}}$ and binding energy $B_{A,Z}$.
In our model, nuclei with $A \geq 2$ are supposed to form an ideal gas, while unbound nucleons are treated within the mean-field theory of nuclear matter (NM).
Interactions among nuclei and with the unbound nucleons are phenomenologically implemented via the excluded volume approximation.
While the assumption that the volume occupied by one nucleus is not available for other nuclei is perfectly consistent with the ideal gas approximation and the classical treatment of nuclei, the case of interactions between nuclei and unbound nucleons is more subtle. 
More precisely, different degrees of permeability of nuclei by the gas of nucleons can be envisaged, which range from totally permeable to totally impermeable~\cite{DinhThi_AA_2023,DinhThi_EPJA_2023}.
The first hypothesis complies with the unbound nucleons being described by plane waves and was adopted in Ref.~\cite{Gulminelli_PRC_2015}.
In contrast, the second hypothesis implicitly assumes that the nucleons are classical particles~\cite{Hempel_2010,Raduta_AA_2025}. 
Its advantage is in treating the nucleons on the same footing with the nuclei.
It is worth mentioning that the excluded volume correction is essential for transitioning to homogeneous matter at densities of the order of $\sat{n}/2$.
As in Ref.~\cite{Raduta_AA_2025}, the full impermeability hypothesis is adopted here.

The key assumption of eNSE consists in the decomposition of the total free energy of the system into a sum of free energies of individual components,
\be
F=\sum_{A,Z} F_{A,Z}+F_{\mathrm{MF}}+F_{\mathrm{el}}+F_{\mathrm{el-el}}^{\mathrm{Coulomb}}.
\label{eq:Ftot}
\ee
Here, $F_{\mathrm{MF}}$ and $F_{\mathrm{el}}$ correspond to the free energies of unbound nucleons and electron gas and $F_{\mathrm{el-el}}^{\mathrm{Coulomb}}$ represents the free energy of the Coulomb interactions between electrons, which can be computed in the Wigner-Seitz approximation~\cite{Lattimer_1985}.
To a very good approximations, $F_{\mathrm{MF}}$ and $F_{\mathrm{el}}$ can be expressed as
\begin{align}
\begin{split}
F_{\mathrm{MF}}(N_n, N_p, T)&= N_n m_n c^2 + N_p m_p c^2 \\
&+ V_{\mathrm{gas}}^{\mathrm{free}} f_{\mathrm{MF}}^0\left(\frac{N_n}{V_{\mathrm{gas}}^{\mathrm{free}}}, \frac{N_p}{V_{\mathrm{gas}}^{\mathrm{free}}}, T \right),
\end{split}
\label{eq:FMF}
\end{align}
and
\be
F_{\mathrm{el}}(N_{\mathrm{el}},T)=V f_{\mathrm{el}}^0(n_{\mathrm{el}},T).
\label{eq:Fe}
\ee
In Eq.~\eqref{eq:FMF}, $m_i$ represents the rest mass of neutrons ($i=n$) and protons ($i=p$),
$N_i$ stands for the total number of unbound nucleons of type $i$ in the volume $V$,
$V_{\mathrm{gas}}^{\mathrm{free}}=(V-V_{\mathrm{cl}})$ is the volume available for these nucleons and
$V_{\mathrm{cl}}$ denotes the volume occupied by clusters and that is not available for the nucleons.
In Eq.~\eqref{eq:Fe}, $f_{\mathrm{el}}^0$ represents the free energy density of an ideal gas of electrons with the density $n_{\mathrm{el}}$.

The first term in Eq.~\eqref{eq:Ftot} corresponds to the free energy of a gas of clusters and is the only ingredient which requires dedicated modeling. In ideal gas approximation, one may write
\ba
F_{A,Z}&=& N_{A,Z} M_{A,Z}^*c^2 \nonumber \\
&-& N_{A,Z} T
\left[ \ln \left( \frac{V_{\mathrm{cl}}^{\mathrm {free}}}{N_{A,Z}} \left(\frac{M_{A,Z}^* T}{2 \pi \hbar^2}\right)^{3/2}\right)
  +1\right],
\label{eq:FAZideal}
\ea
where $N_{A,Z}$ denotes the number of the $(A, Z)$-nuclei, $M_{A,Z}^*$ stands for their mass and $V_{\mathrm{cl}}^{\mathrm {free}}=V-V_{\mathrm {gas}}-V_{\mathrm{cl}}$ represents the volume available for the translational movement of nuclei after the volume occupied by unbound nucleons $V_{\mathrm {gas}}=(N_n+N_p) v_0$ and all nuclear species $V_{\mathrm{cl}}=\sum_{A,Z} N_{A,Z} v_{A,Z}$ is subtracted.
It is pretty obvious that, in addition to $A$ and $Z$-numbers, $M_{A,Z}^*$ also depends on temperature, density of unbound nucleons, Coulomb interactions among the protons in the nucleus and Coulomb interactions between the protons in the nucleus and the electrons in the surrounding gas.

\looseness=-1
Temperature induced modifications of nuclear radii, binding energies and level densities of intermediate mass and heavy nuclei along with the issue of equilibrium with a surrounding gas of nucleons have been addressed within the Hartree-Fock~\cite{Bonche_NPA_1984,Bonche_NPA_1985,Vautherin_PLB_1987,Papakonstantinou_PRC_2013} and Thomas-Fermi~\cite{De_PRC_2012} approaches of nuclear matter.
These studies indicate that for temperatures in excess of a certain $(A,Z)$- and interaction-dependent value nuclei cease to exist.
This phenomenology can be roughly accounted for via the liquid-drop model, which allows to express the free energy of a nucleus as an analytical function of its density and temperature ~\cite{Randrup_PRC_1992,Parmar_PRC_2022}.
The case of light nuclei, for which the mean field approximation does not apply, was addressed via a microscopic quantum statistical approach~\cite{Typel_PRC_2010,Ropke_PRC_2011}. 
Results corresponding to $^2$H, $^3$H, $^3$He and $^4$He embedded in a symmetric nucleon gas show that the binding energy of these nuclei decreases with the density of the medium and that the higher the temperature the lower the rate of this decrease~\cite{Typel_PRC_2010}.
It is reasonable to assume that the same holds for other light nuclei embedded in a nucleon gas of arbitrary proton fraction, but to the best of our knowledge no such calculation is available to date.

The inadequacy of the liquid drop model to account for microscopic effects like pairing and shell closure makes that, for the relatively low temperatures of neo-NSs, the use of experimental nuclear masses represents the best compromise for the description of cluster energetics.
Similarly, in default of exhaustive microscopic information regarding in-medium modifications of light clusters, we choose to use the vacuum values. 
With these simplifications, it is clear that the only modifications of the clusters' properties enter via the excluded volume correction.
According to Refs.~\cite{Hempel_PRC_2011,Typel_2016}, with the exception of densities slightly lower than the density where a species disappears, the excluded volume correction is a fair surrogate for the Pauli blocking of clusters states by the nucleons in the medium.

The most tractable version of this model corresponds to the approximation
\be
F_{A,Z}=F_{A,Z}^{\mathrm{vacuum}}+F_{A,Z}^{\mathrm{int}}+F_{A,Z}^{\mathrm{Coulomb}},
\ee
where $F_{A,Z}^{\mathrm{vacuum}}$ is computed as in Eq.~\eqref{eq:FAZideal} but assuming that $M_{A,Z}^*=M_{A,Z}$,
$F_{A,Z}^{\mathrm{int}}$ accounts for population of excited states, made possible by finite temperatures,
\be
F_{A,Z}^{\mathrm{int}}=-T N_{A,Z} \ln \tilde z_{A,Z}^{\mathrm{int}},
\label{eq:FAZ_int}
\ee
and the Coulomb contribution to the free energy is
\be
F_{A,Z}^{\mathrm{Coulomb}}=N_{A,Z} E_{A,Z}^{\mathrm{Coulomb}}.
\label{eq:FAZCoul}
\ee
The internal partition function entering Eq.~\eqref{eq:FAZ_int} is given by
\be
\tilde z_{A,Z}^{\mathrm{int}}=g_\mathrm{G.S.}+\int_0^{B(A,Z)} d \epsilon \exp(-\epsilon/T) \rho(\epsilon),
\ee
where $g_\mathrm{G.S.}$ stands for the spin degeneracy of the ground state, $B(A,Z)$ is the binding energy and $\rho(\epsilon)$ represents the level density.
In the Wigner-Seitz approximation, $E_{A,Z}^{\mathrm{Coulomb}}$ writes~\cite{Lattimer_1985}:
\be
E_{A,Z}^{\mathrm{Coul}}=-\frac35 \frac{Z^2 e^2}{R_{A,Z}} \left( \frac32 x - \frac12 x^3\right),
\label{eq:ECoul}
\ee
where $x=\left[A/Z \cdot n_{\mathrm{el}}/\sat{n}(\delta) \right]^{1/3}$, $\delta=1-2Z/A$,
$R_{A,Z}=\left[3 v_{A,Z}/4\pi \right]^{1/3}$,
$n_{\mathrm{el}}=\left[N_p + \sum_{A,Z} Z N_{A,Z}\right]/V$.
Note that Eq.~\eqref{eq:ECoul} accounts for Coulomb interactions between protons and electrons and among the electrons.
Inclusion of electron-electron interactions in $E_{A,Z}^{\mathrm{Coul}}$ renders the term $F_{\mathrm{el-el}}$ in Eq.~\eqref{eq:Ftot} unnecessary.
For a detailed description of our extended NSE model, see Ref.~\cite{Raduta_AA_2025}.
A similar model was proposed in Ref.~\cite{Hempel_2010}.

The robustness of our results was tested by alternatively considering different pools of nuclei:
i) nuclides present in the experimental mass table AME~2020~\cite{AME2020} supplemented by nuclides in the DZ10 table by Duflo and Zuker~\cite{Duflo_PRC_1995} (reference set),
ii) the same as above but with an extended DZ10 table (T1 test set),
iii) nuclei in the AME~2020 table only (T2 test set).
The extended DZ10 table was generated by extending the isotopic chains of nuclei with $Z \geq 11$ both in the neutron-rich and proton-rich sides. Nuclear masses were computed with the code \texttt{du\_zu\_10.feb96fort}~\footnote{Available at \url{https://www-nds.iaea.org/amdc/Duflo-Zuker-program.zip}} for nuclei with $(1-f) A_i(Z) \leq A < A_i(Z)$ and $A_s(Z) < A \leq A_s(Z) (1+f)$, where $A_i(Z)$ and $A_s(Z)$ are the lighest and heaviest $Z$-isotopes in the DZ10 table and $f=0.3$.
For i) and ii), whenever available, experimental values are preferred.
Thermal population of excited states is allowed only for nuclei with $A \geq 16$. For the level density, the back-shifted Fermi gas parametrization from Refs.~\cite{Bucurescu_2005,vonEgidy_2005} is employed.

For unbound nucleons we employ the non-relativistic model of NM~\cite{Negele_1972,Vautherin_1996}, but all important features would persist should this formalism be replaced with the covariant density functional theory of dense matter.
In particular, BSk25~\cite{BSk22-BSk26} and BBSk1~\cite{Raduta_AA_2025} versions of the Brussels extended Skyrme effective nucleon-nucleon interaction were used. 

The beta equilibrium configurations were computed within the zero-temperature approximation by requiring that $\mu_n = \mu_p + \mu_e$, which corresponds to vanishing lepton chemical potential. 
As discussed in Ref.~\cite{AH_2018}, this is a valid approximation as long as  $T < 1$~MeV. 
Thus, we only consider the lower end of the neo-NS temperature range in our investigations.
The results presented in the next section correspond to $T = 0.4$~MeV and $T = 0.1$~MeV.
Comparison of eNSE results at $T=0.1~\mathrm{MeV}$ with zero-temperature calculations of NS crusts in the literature highlight the contributions of translational degrees of freedom, which exist only at $T>0$.

\section{Crust composition}
\label{sec:Results}

\begin{figure}
	\includegraphics[scale=0.49,trim={0.2cm 0.3cm 2.7cm 1.1cm}, clip]{"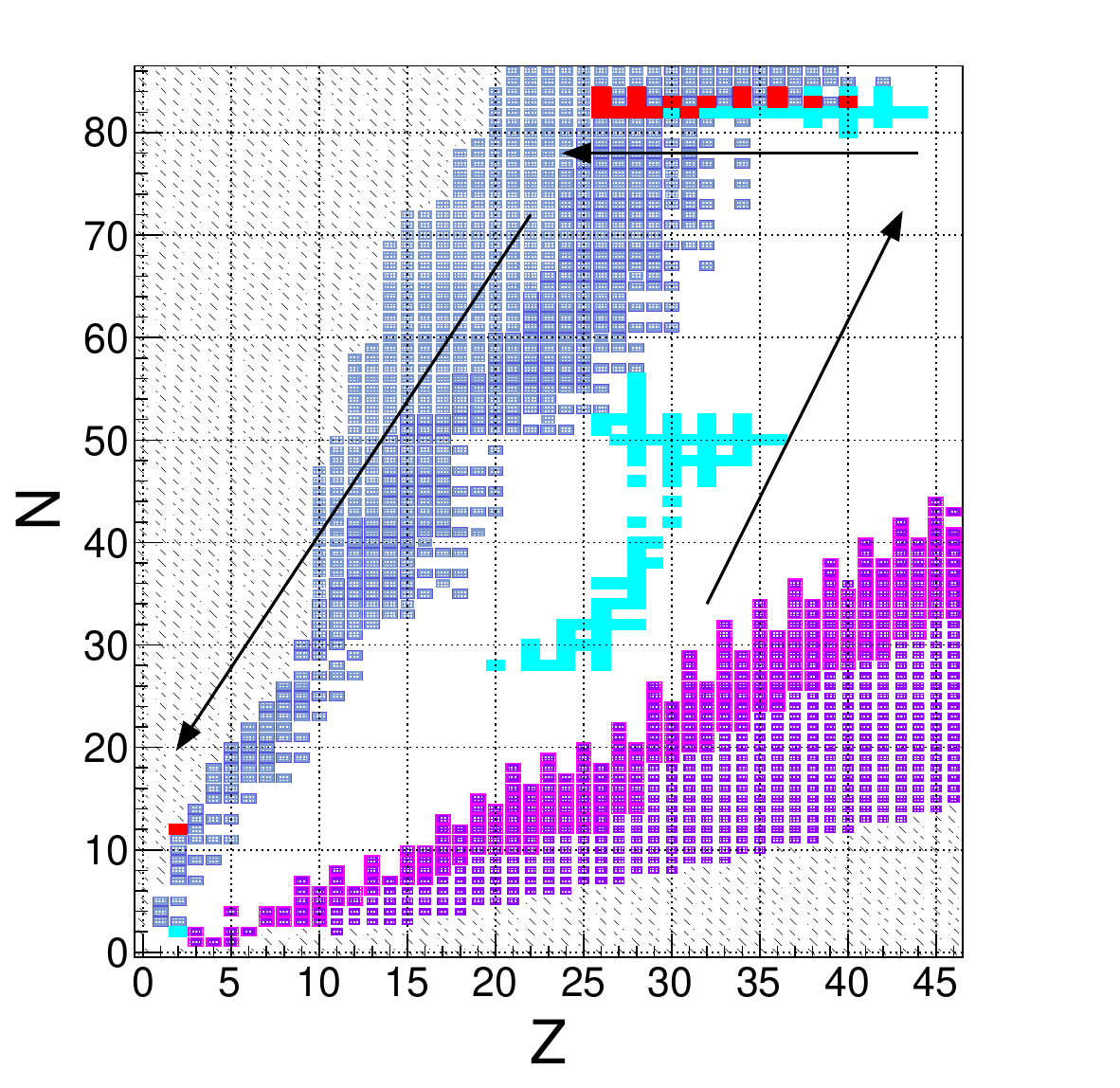"}
	\caption{Composition of the neo-NS crust and visualization of the employed mass tables.
	White areas represent the nuclides with positive neutron and proton separation energies, while the
    nuclides with negative neutron and proton separation energies are marked in violet and magenta, respectively.
    Light (dark) hues are used for the T1 (reference) set.
	The gray hatched domain corresponds to nuclei for which nuclear mass data are not available in T1.
	Cyan and red show the ``stable'' and beyond neutron-drip nuclei present in the crust.
	The arrows show the direction of increase of $n_\mathrm{B}$.
	The results correspond to T1 run, $T=0.4$~MeV and BSk25~\cite{BSk22-BSk26} effective interaction. 
	}
	\label{Fig:Nuclides}
\end{figure}

\begin{figure}
\includegraphics[scale=0.435,trim={0.1cm 0.1cm 0.1cm 0.1cm}, clip]{"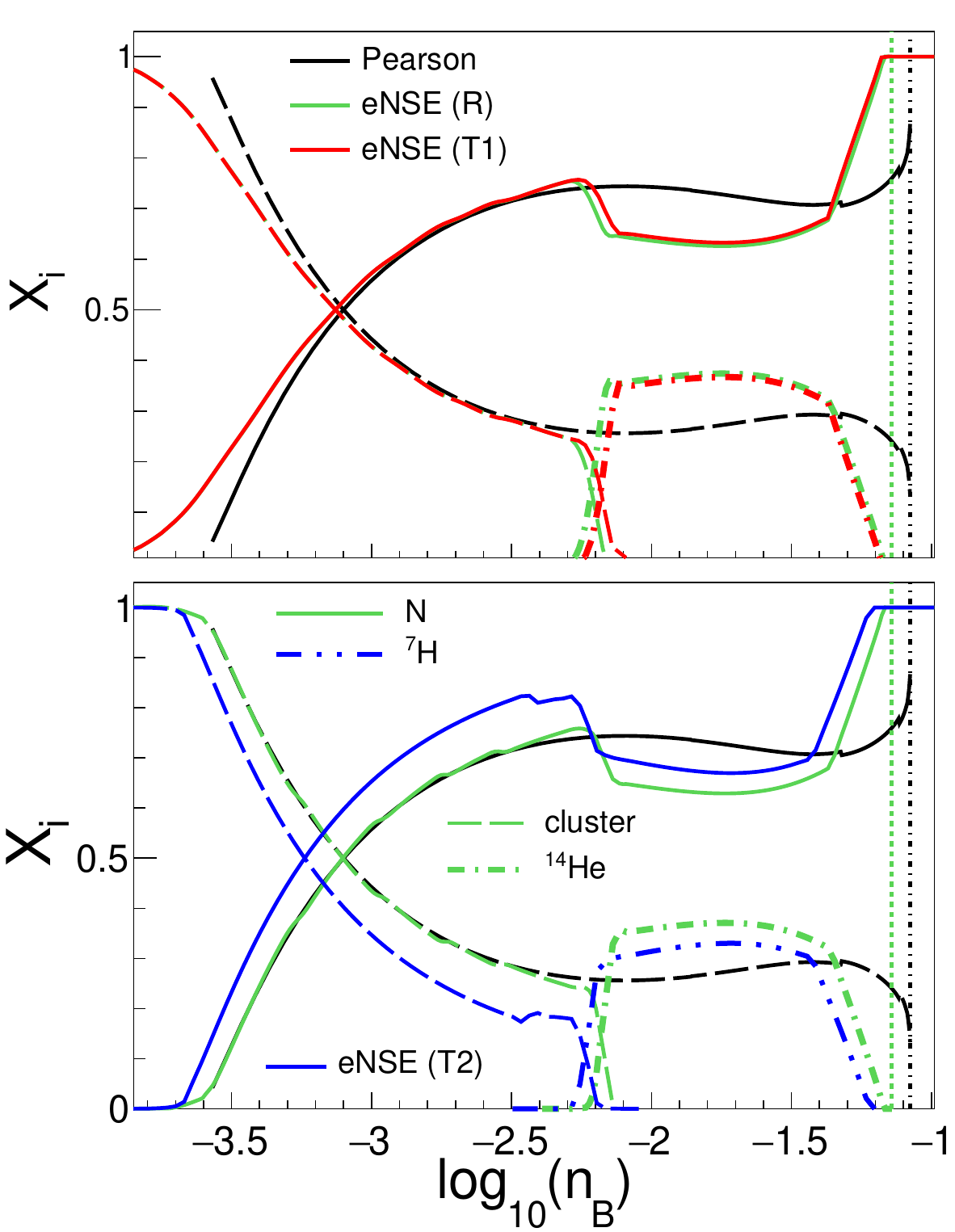"}
\caption{Mass fractions of ``heavy'' ($A \geq 20$) nuclei, light neutron-rich nuclei $^{14}$He (for the reference and T1 runs) and $^{7}$H (for T2 run), and unbound nucleons as functions of $n_\mathrm{B}$ for $\beta$-equilibrated crust at $T=0.1~\mathrm{MeV}$ (bottom) and  $T=0.4~\mathrm{MeV}$ (top).
eNSE results are compared with the zero temperature results of Ref.~\cite{Pearson_MNRAS_2018}. 
In all cases, the BSk25~\cite{BSk22-BSk26} effective interaction was employed.
Vertical dotted lines mark the transition to homogeneous matter, $n_{\mathrm{tr}} = 0.0742~\mathrm{fm}^{-3}$ and $0.0839~\mathrm{fm}^{-3}$ for eNSE reference run and the results of Ref.~\cite{Pearson_MNRAS_2018}, respectively.
}
\label{Fig:MassFrac}
\end{figure}

Let us start with the composition of the crust.
The most interesting and novel result here is the rapid change of composition from heavy nuclei with $A \sim 100$ to an almost pure layer of neutron-rich light nuclei in the innermost shells of neo-NS crusts. Specifically, those are $^{14}$He ($^7$H) for the reference and T1 test (T2 test) runs, see Figs.~\ref{Fig:Nuclides} and \ref{Fig:MassFrac}.
This layer persists for temperatures ranging from 0.1 MeV, which is beyond the validity limit of NSE, to several MeV. 
$^{14}$He stems from the DZ10 table~\cite{Duflo_PRC_1995}; it has a binding energy of 19.3~MeV and a negative neutron separation energy, $S_n=-1.29~\mathrm{MeV}$.
$^7$H is present in the AME2020 table; its binding energy and neutron separation energy are 6.58~MeV and $S_n=0.81~\mathrm{MeV}$ , respectively.

Figure~\ref{Fig:Nuclides} provides the composition of the crust for the T1 run at $T=0.4~\mathrm{MeV}$.
Nuclides with positive neutron and proton separation energies are marked in cyan; nuclides beyond neutron-drip are marked in red.
With the exception of ($Z=26$, $N=83$) and ($Z=26$, $N=84$), the reference set provides the same composition.
The arrows give a general idea of the direction of the increase in density.
The preference for nuclides with magic number of neutrons (especially 50 and 82) and protons (28) is clearly visible. 
Some features of the employed nuclear mass tables are displayed as well, see the figure caption.

Further insight into crust composition is provided in Fig.~\ref{Fig:MassFrac}, which illustrates
the dependence of the mass fractions of ``heavy'' ($A \geq 20$) nuclei, $^{14}$He (for reference and T1 runs) or $^{7}$H (for T2 run), and unbound nucleons on the total baryon number density $n_\mathrm{B}$ for $T=0.1$~MeV (bottom panel) and $T=0.4$~MeV (top panel).
The predictions of the extended Thomas Fermi plus Strutinsky integral model of Ref.~\cite{Pearson_MNRAS_2018}, which corresponds to $T=0$, are shown as well for comparison (black curves).
As before, the results correspond to BSk25~\cite{BSk22-BSk26}.
The transition density to homogeneous matter for our eNSE model at the lowest temperature and the model of Ref.~\cite{Pearson_MNRAS_2018} are indicated with vertical dotted lines.
Figure~\ref{Fig:MassFrac} demonstrates that, for $n_\mathrm{B} \lesssim 0.006$~fm$^{-3}$ and $T=0.1$~MeV, the reference run provides the same mass fractions as the model of Ref.~\cite{Pearson_MNRAS_2018}; the same holds for the T1 run (not shown);
for other effective interactions (not shown) the agreement is also very good.
When the pool of nuclei is restricted to AME2020 table, the eNSE results deviate from those of Ref.~\cite{Pearson_MNRAS_2018}.
The fraction of mass bound in clusters decreases and, due to mass conservation, more unbound nucleons exist.
The agreement between our results for the reference and T1 runs and those in Ref.~\cite{Pearson_MNRAS_2018} suggests that, as long as sufficiently neutron-rich nuclei are allowed to nucleate, the mass sharing is mainly determined by the effective interaction.
This situation can be seen as an a posteriori justification for using eNSE beyond its validity limit.

From Fig.~\ref{Fig:MassFrac} one can see that, for both temperatures, the $^{14}$He $(^{7}\mathrm{H})$ layer rapidly becomes abundant at $n_\mathrm{B} \approx 0.006$~fm$^{-3}$ and extends up to the transition density. 
It comprises almost 40\% (30\%) of the mass and is responsible for the depletion of the gas of unbound nucleons, as well as a sudden increase of the $\langle A_{\mathrm{cl}}\rangle/\langle Z_{\mathrm{cl}} \rangle$ ratio to $7$, see below.
The onset density and the mass fraction depend on the effective interaction and also, in principle, on the temperature;
the lowest onset density and the highest mass fraction among all the interactions that we have tested correspond to BSk25, which has high values of the symmetry energy for $n < \sat{n}$.
The emergence of the $^{14}$He $(^{7}\mathrm{H})$ layer causes our finite temperature results to depart from the $T=0$ results of Ref.~\cite{Pearson_MNRAS_2018}. 
In principle, two explanations can be put forward for this situation.
First, one can suspect that this is the outcome of the pool limitation. The fact that, with the exception of $^{109}$Fe and $^{110}$Fe that are present over $5.75 \cdot 10^{-3}~\mathrm{fm}^{-3} < n < 6.92 \cdot 10^{-3}~\mathrm{fm}^{-3}$, the same crust composition is obtained for the reference and T1 runs dismisses this hypothesis. Alternatively,
this replacement of more massive nuclei by neutron-rich light nuclei in the deepest parts of the inner crust can be explained as the effect of translational degrees of freedom, which do (do not) exist at finite (zero) temperature.
This mechanism was previously noticed by Dinh Thi et al.~\cite{DinhThi_AA_2023,DinhThi_EPJA_2023}, who allowed for various permeability degrees of nuclear clusters with respect to unbound nucleons. The cluster definition used here corresponds to
the zero-permeability of Refs.~\cite{DinhThi_AA_2023,DinhThi_EPJA_2023}, situation in which the light clusters abundance is the highest among all scenarios analyzed by Dinh Thi et al. The qualitative agreement between our results and those in  Refs.~\cite{DinhThi_AA_2023,DinhThi_EPJA_2023} show that presence of light nuclei in the deep crust is not an artefact of our model but rather a real possibility. It remains to check that this composition survives the competition with the ``nuclear pasta'' phases and implementation of a more realistic treatment of in-medium modifications of cluster energetics.

The high purity of the $^{14}$He $(^{7}\mathrm{H})$ layer may lead to very different thermal and electrical conductivities compared to layers made of massive nuclei. 
Indeed, in our case the impurity parameter $Q_\mathrm{imp}$~\cite{PC_2021,SS_18} is almost density independent and very low ($Q_\mathrm{imp} \lesssim 0.01$) in the whole region where $^{14}$He is dominant and $T < 0.5$~MeV. 
In contrast, in Ref.~\cite{PC_2021} the impurity parameter oscillates a lot as a function of $n_\mathrm{B}$ in the range $0.01 \lesssim Q_\mathrm{imp} \lesssim 100$ for $0.01 \leq n_\mathrm{B} \leq 0.08$~fm$^{-3}$ and $T \approx 0.27$~MeV with the amplitude and position of the extrema depending on the effective interaction. This feature will affect the interpretation of crustal cooling in low-mass X-ray binaries~\footnote{Notice, however, that our results refer to non-accreted crusts, while low mass X-ray binaries have partially or fully accreted crusts.} (see, e.g., Refs.~\cite{PCC_2019,PC_2021,Suleiman_2022}), cooling of young isolated NSs during the crust relaxation phase as well as the magneto-rotational evolution of NSs (see, e.g., Ref.~\cite{Pons_2013}).

Then, $^{14}$He $(^{7}\mathrm{H})$ will alter the process of crust crystallization, which is essential for the crust structure, composition, and transport properties. 
We remind that the crust crystallizes as a Coulomb crystal and, in the simplest one-component plasma approximation, the crystallization temperature is proportional to $Z^2/A_\mathrm{cell}^{1/3}$~\cite{Chamel_2008}.
Thus, a sudden change of $Z$ from $20 \leq Z \leq 40$  to $Z = 1,2$ will diminish the crystallization temperature by a factor of a few hundreds (a change of $A_\mathrm{cell}$ is of little importance due to the power $1/3$). 
Moreover, $^{14}$He $(^{7}\mathrm{H})$ leads to a configuration where more dilute shallower layers of the crust might already be frozen, while the deeper and denser layers still remain liquid.
Also important is the ion plasma temperature, which delineates the classical and quantum plasma regimes and is proportional to $Z$~\cite{SS_18}.
Again, the onset of $^{14}$He $(^{7}\mathrm{H})$ will cause a sudden change in this temperature.

Since we employed experimental and evaluated mass tables, all our results assume implicitly that in dense media all nuclei, including the light species, have the same properties as in vacuum. 
Although this is certainly not true, the studies available to date~\cite{Typel_PRC_2010,Papakonstantinou_PRC_2013} are not sufficient for a realistic treatment of nuclear binding energies over the whole mass table and for arbitrary densities and isospin compositions of the medium.
From our reference and test runs as well as calculations in Refs.~\cite{DinhThi_AA_2023,DinhThi_EPJA_2023}, it is clear that the presence of a particular exotic light cluster depends on the employed pool of nuclei.
However, the overall tendency to replace heavy nuclei with light ones deep in the inner crust seems to be a robust conclusion. 
Further investigation of this issue is necessary, possibly accounting for other exotic light clusters that may be relevant for the crust, e.g., tetraneutron resonances~\cite{Ivanytskyi_2019,Pais_2023}.

Nuclear burning of light elements deep in the inner crust has to be investigated further as well. 
As long as one deals with neo-NSs, this is not an issue because the NSE holds, which means that the nuclear reaction rates are irrelevant and the burning is compensated by photo-disintegration reactions.
However, as the neo-NS cools down, the system moves out of the NSE and the reaction rates become important.
Unfortunately, they are very uncertain for the conditions of the inner crust.
One can speculate that the impact of thermonuclear reactions will be determined by the competition between the burning and cooling timescales.
Pycnonuclear burning, being temperature-independent, may completely destroy the light elements in the inner crust.
As such, the question of the composition of the inner crusts of cold isolated NSs remains open, but it is outside the scope of the present paper.

\section{Stability with respect to diffusion}

\begin{figure}
\includegraphics[scale=0.435,trim={0.1cm 0.1cm 0.1cm 0.1cm}, clip]{"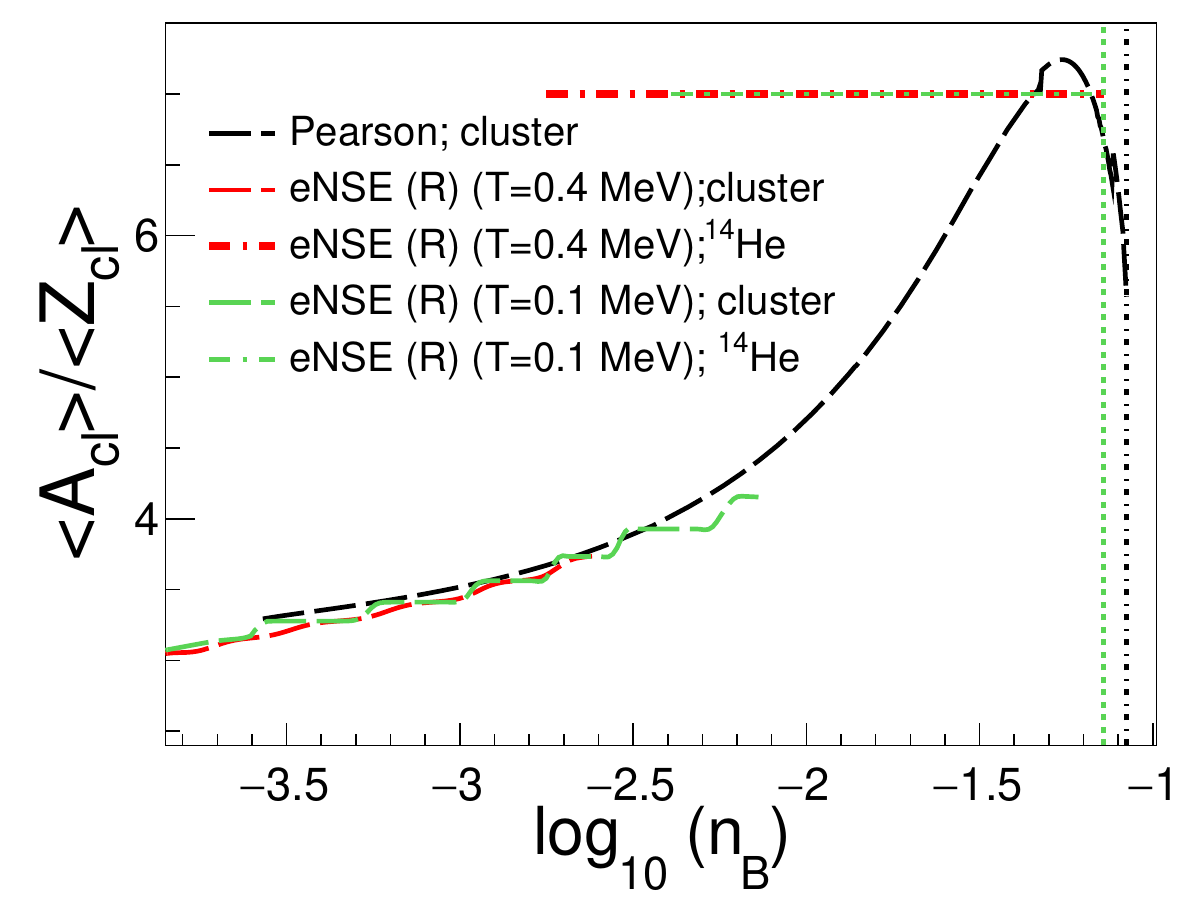"}
\caption{Average mass to average charge numbers ratio as a function of $n_\mathrm{B}$.
eNSE results for beta-equilibrated matter are compared with the zero temperature results of Ref.~\cite{Pearson_MNRAS_2018}. 
Results corresponding to BSk25~\cite{BSk22-BSk26}.
As in Fig.~\ref{Fig:MassFrac}, vertical dotted lines mark the transition to homogeneous matter.
}
\label{Fig:Buoy}
\end{figure}

The next aspect that we want to discuss is the problem of buoyancy and the (in)stability of the inner crust with respect to diffusion of ions.
For an electrically neutral multicomponent plasma in an external gravitational field, the velocity of gravitational separation (sedimentation) of the ions is determined by the ``effective weight'' $A/Z$ (see Appendix~\ref{App:Diff} and Refs.~\cite{CBA_2010,BY_2013}).
As long as the $A/Z$ ratio is a non-decreasing function of $n_\mathrm{B}$, the configuration is stable, since ions with larger $A/Z$ ratio are below ions with smaller $A/Z$ ratio.
However, if the ``effective weight'' decreases with $n_\mathrm{B}$, the situation becomes unstable.
This is similar to the classical Rayleigh–Taylor instability, only here the buoyancy is determined by the $A/Z$ ratio.

Fig.~\ref{Fig:Buoy} depicts the dependence of the $\langle A_{\mathrm{cl}}\rangle/\langle Z_{\mathrm{cl}}\rangle$ ratio on $n_\mathrm{B}$ for $T=0.1$ and $0.4~\mathrm{MeV}$, as obtained in the reference run. Quite remarkably, the $\langle A_{\mathrm{cl}} \rangle/\langle Z_{\mathrm{cl}} \rangle$ ratios are also very similar for the reference run and Ref.~\cite{Pearson_MNRAS_2018}, 
even though the values of $\langle A_{\mathrm{cl}}\rangle$ and $\langle Z_{\mathrm{cl}} \rangle$ predicted by these two models are very different. More precisely, Ref.~\cite{Pearson_MNRAS_2018} allows for much more massive and much more neutron-rich clusters than eNSE.
The ``steps'' caused by the shell and odd-even effects coming from the employed experimental and evaluated mass tables are very clearly seen in the top panel at $T = 0.1$~MeV (green dashed curve). 
One can also see how these ``steps'' are partially smoothed out by temperature (red dashed curve).
It turns out that our eNSE model is stable with respect to diffusion, and neutron-rich light nuclei play a big role in this aspect.
The model of Ref.~\cite{Pearson_MNRAS_2018} is not stable with respect to diffusion, as it has a range of densities where the $A_{\mathrm{cl}}/Z_{\mathrm{cl}}$ ratio decreases. 
We have also checked some other existing crustal EOSs.
All EOSs in Ref.~\cite{Pearson_MNRAS_2018} as well as GPPVA EOSs~\cite{Grill_2014} are unstable.
On the other hand, NV EOS~\cite{Negele_1973}, D1M and D1M$^*$ EOSs~\cite{Vinas_2021}, and SPG EOSs~\cite{Scurto_2024} are stable.

This buoyancy issue is different from the issue of the (in)stability of inner crusts with respect to diffusion of unbound neutrons~\cite{Chugunov_2020,GC_2020}.
Unlike the latter, the former seems to be of little practical significance.
When the crust crystallizes, ion diffusion becomes strongly suppressed and the buoyancy instability becomes much less relevant.
When the crust is warm enough to be liquid, the NSE likely still holds. 
In that case, nuclear reactions are much faster than ion diffusion and adjust very quickly any changes in composition introduced by diffusion.

\section{Thermodynamic stability}

\begin{figure}
	\includegraphics[scale=0.43]{"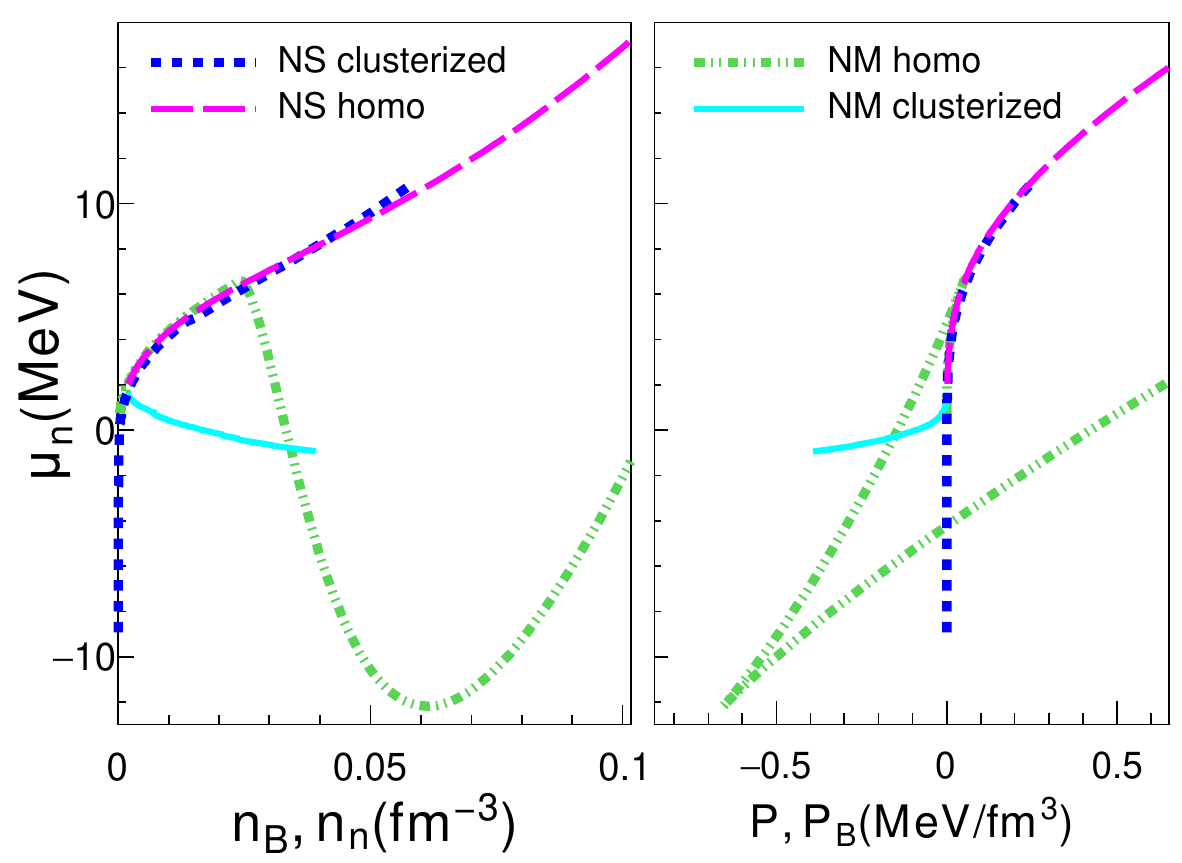"}
	\caption{Thermodynamic (in)stability of cold NS matter ($T=0.1$~MeV).
	Left and right panels show the baryon chemical potential ($\mu_\mathrm{B}=\mu_n$) as a function of particle number density ($n_\mathrm{B}$ or $n_n$) and pressure ($P$ or $P_\mathrm{B}$), respectively.
	NM without Coulomb interaction is computed at a fixed $\mu_p=-30$~MeV.
	NS matter with electrons is computed at a fixed $\mu_\mathrm{lep} = 0$ (beta equilibrium).		 
	BBSk1~\cite{Raduta_AA_2025} effective interaction.
	See text for details.
	}
	\label{Fig:Stability}
\end{figure}

The last question that we want to discuss is the question of thermodynamic stability of clusterized NS matter.
Since the domain of density-driven instabilities in dilute homogeneous NM shrinks with increasing temperature~\cite{Mueller_1995,Ducoin_2006,Vidana_2008,Carbone_2018}, it is sufficient to analyze the case of the lowest considered temperature, $T=0.1$~MeV.

To investigate the question of thermodynamic stability of a multicomponent system, one may plot the chemical potential of one component as a function of its conjugate number density while keeping the chemical potential of the other components fixed~\cite{Ducoin_2006,Ducoin_2007}.
If there is a range of densities where the chemical potential decreases with increasing density, then the system has an instability with respect to phase separation. 
In Fig.~\ref{Fig:Stability} we compare four systems: clusterized (cyan) and homogeneous (green) NM without Coulomb interaction~\footnote{To be more precise, for inhomogeneous matter the electron contribution was subtracted, but the Coulomb contribution remained as if the electrons were present.} and clusterized (blue) and homogeneous (magenta) NS matter (i.e., with electrons and in beta equilibrium). 
For NM we plot $\mu_n$ as a function of $n_n$ (left panel) at a fixed $\mu_p =-30$~MeV; rest-mass contributions are subtracted.
For NS matter, we plot the baryon chemical potential $\mu_\mathrm{B}(=\mu_n)$ as a function of $n_\mathrm{B}$ (left panel) for a fixed lepton chemical potential $\mu_\mathrm{lep} = 0$ (beta equilibrium).
From the left panel, one can immediately see that both clusterized and homogeneous NM have regions where $\mu_n$ decreases with $n_n$ and, thus, both are thermodynamically unstable with respect to the first-order phase transition. 
However, for both clusterized and homogeneous NS matter, $\mu_\mathrm{B}$ is a monotonically increasing function of $n_\mathrm{B}$.
Such matter is thermodynamically stable. 
This result shows that the instabilities with respect to finite density fluctuations that manifest in NM with electrons~\cite{Baym_NPA_1971,Pethick_NPA_1995,Ducoin_2007} are completely exhausted by clusterization.
The right panel provides further information in terms of $\mu_n$ vs $P_\mathrm{B}$ (baryonic pressure) for NM at fixed $\mu_p = -30$~MeV and $\mu_\mathrm{B}(=\mu_n)$ vs $P$ (total pressure) for NS matter at fixed $\mu_\mathrm{lep} = 0$.
Over the phase instability domains, the $\mu_n(P_\mathrm{B})$ curves (green and cyan) are concave.
Whenever the system is stable, these curves are convex.

In conclusion, there is no thermodynamic instability; the clusterized stellar matter does transform into the homogeneous one by dissolution or disintegration of clusters. 
The physical mechanism of disintegration of clusters is mentioned, e.g., in Ref.~\cite{GC_2020} in the context of accreted crusts. 
The mechanism of dissolution is discussed in Ref.~\cite{Pais_WS_2017}.

\section{Conclusions}
\label{sec:Concl}

An extended NSE model~\cite{Raduta_AA_2025} was employed here to compute the composition of subsaturated catalyzed nuclear matter with temperatures of $T \approx 0.4~\mathrm{MeV}$, relevant for the crusts of neo-NSs.
Interactions between nuclei as well as between nuclei and unbound nucleons are treated within the excluded volume paradigm, which is a fair approximation up to densities slightly lower than the density where the nuclei dissolve into the gas of unbound nucleons.
More accurate calculations would require systematic implementation of microscopically estimated binding energy shifts, which are not available yet.

A possible presence of an almost pure layer of neutron-rich light nuclei in the innermost shells of the crusts of neo-NSs was pointed out. 
Tests with different mass tables show that the exact composition depends on the pool of eNSE nuclei. Results in Ref.~\cite{DinhThi_AA_2023,DinhThi_EPJA_2023} additionally show that the composition depends on the degree of permeability of nuclei with respect to the gas of unbound nucleons, with the hypothesis of complete impenetrability predicting the largest amounts of light clusters.
The deviation of finite temperature results from results at $T=0$ is most probably due to the translational degrees of freedom of nuclei, which only exists at $T>0$.
Even if a definite answer regarding this issue commends implementation of microscopically derived in-medium modifications of clusters' binding energy, it is clear that warm NSs' crusts will have transport and elastic properties different than those of cold NSs' crusts. 
Also, it remains to demonstrate whether light exotic species in the deep crust survive the competition with deformed macroscopic structures should the latter be allowed to form.

In addition, we showed that, because of light nuclei, warm NSs' crusts are stable with respect to the diffusion of ions (buoyancy), which does not seem to be always the case for other models of the crusts. 
Again, a definite answer requires more sophisticated investigations, which are beyond our present scope. Finally, we demonstrated that clusterization completely exhausts the thermodynamic instabilities of charge neutral mixtures of nucleons and electrons.

\begin{acknowledgments}	
	We express our gratitude to the anonymous referees for all their comments and suggestions that helped improving our manuscript.
	The authors acknowledge support from a grant from the Ministry of Research, Innovation and Digitization, CNCS/CCCDI–UEFISCDI, Project No. PN-IV-P1-PCE-2023-0324 and partial support from Project No. PN 23 21 01 02.
	
	M.V.B. and A.R.R. contributed equally to this work.
\end{acknowledgments}

\appendix

\section{Diffusion in the inner crust}
\label{App:Diff}
This derivation of the diffusive currents closely follows Ref.~\cite{BY_2013} with modifications appropriate for the inner crust, namely the inclusion of unbound nucleons (see also Ref.~\cite{Chugunov_2020}). 
The temperature is assumed to be constant, the effects of general relativity are neglected (thus, there is no difference between local and redshifted temperatures) and the crust is considered in the plane-parallel approximation.

The composition is: ions of type 1 and 2, unbound nucleons, and electrons.
Neutron and proton superfluidity is disregarded.
We can write the generalized forces as follows:
\begin{align}
	&\mathbf{F}_j = Z_j e \mathbf{E} + A_j m_0 \mathbf{g} - \mathbf{\nabla} \mu_j, \quad j = 1, 2,
	\label{eq:F_j} \\
	&\mathbf{F}_n = m_0 \mathbf{g} - \mathbf{\nabla} \mu_n, 
	\label{eq:F_n} \\
	&\mathbf{F}_p = e \mathbf{E} + m_0\mathbf{g} - \mathbf{\nabla} \mu_p, 
	\label{eq:F_p} \\
	&\mathbf{F}_e = - e \mathbf{E} - \mathbf{\nabla} \mu_e. 
	\label{eq:F_e}
\end{align}
Here $\mathbf{g}$ is the free-fall acceleration (assumed to be constant), $e$ is the elementary charge, $\mathbf{E}$ is the electric field that maintains charge neutrality ($n_e = Z_1 n_1 + Z_2 n_2 + n_p$). 
In this derivation, chemical potentials do \emph{not} include rest-masses and potentials of external forces (i.e., of gravity force and electric field).
As discussed in Ref.~\cite{CBA_2010}, the mass defect of nuclei has a negligible impact on the diffusion of ions and thus can be neglected. Similarly, we have neglected the difference between the masses of a proton and a neutron ($m_p \approx m_n = m_0$) and the electron mass in Eq.~\eqref{eq:F_e}.

The hydrostatic equilibrium of the whole system can be written as 
\begin{align}
	\mathbf{\nabla} P = \rho \mathbf{g} 
	\label{eq:hydro}
\end{align}
with $P$ being the total pressure, $\rho \approx m_0(A_1 n_1 + A_2 n_2 + n_n + n_p)$ being the total mass density, and the contribution of the electrons to the mass density is marginal.
The deviation of species $\alpha$ from mechanical equilibrium is
\begin{align}
	\mathbf{d}_\alpha = \frac{m_\alpha n_\alpha}{\rho}\sum_\beta n_\beta\mathbf{F}_\beta \ -n_\alpha \mathbf{F}_\alpha = -n_\alpha \mathbf{F}_\alpha, \quad \alpha = 1, 2, e, n, p,
	\label{eq:d_a}
\end{align}
where the first term is zero due to the overall hydrostatic equilibrium: $\sum_\beta n_\beta\mathbf{F}_\beta = \mathbf{\nabla} P - \rho \mathbf{g} = 0$~\cite{BY_2013}. From the definition above, it is obvious that $\sum_\alpha \mathbf{d}_\alpha = 0$.
The diffusion current is
\begin{align}
	\mathbf{J}_\alpha = \Phi \sum_{\beta \neq \alpha} m_\alpha m_\beta D_{\alpha \beta} \mathbf{d}_\beta
	\label{eq:J}
\end{align}
with $D_\mathit{\alpha \beta}$ being the (generalized) diffusion coefficients and $\Phi$ being a normalization factor, which is not important here.

Since electrons are very light, they adjust almost instantaneously to any mechanical changes in the baryon subsystem.
In other words, electrons are always in diffusive equilibrium, and we can factor them out of a problem of ion diffusion.
This is similar to the Born-Oppenheimer approximation.
Thus, 
\begin{align}
	\mathbf{d}_e = 0, \qquad - e \mathbf{E} - \mathbf{\nabla} \mu_e = 0.
	\label{eq:e_equil}
\end{align}

For ions plus unbound nucleons, we have ($m_j \approx A_j m_0$):
\begin{align}
	\mathbf{J}_1 = \Phi \left(m_1 m_2 D_{12}\mathbf{d}_2 + m_1 m_0 D_{1n} \mathbf{d_n} + m_1 m_0 D_{1p} \mathbf{d_p} \right)
	\label{eq:J_1} \\
	\mathbf{J}_2 = \Phi \left(m_1 m_2 D_{21}\mathbf{d}_1 + m_2 m_0 D_{2n} \mathbf{d_n}  + m_2 m_0 D_{2p} \mathbf{d_p}\right)
	\label{eq:J_2}
\end{align}
with
\begin{align}
	&\mathbf{d}_1 = -n_1 Z_1 e \mathbf{E} - n_1 A_1 m_0 \mathbf{g} + n_1 \mathbf{\nabla} \mu_1, 
	\label{eq:d_1} \\
	&\mathbf{d}_2 = -n_2 Z_2 e \mathbf{E} - n_2 A_2 m_0 \mathbf{g} + n_2 \mathbf{\nabla} \mu_2,
	\label{eq:d_2} \\
	&\mathbf{d}_n = -n_n m_0 \mathbf{g} + n_n \mathbf{\nabla} \mu_n,
	\label{eq:d_n} \\
	&\mathbf{d}_p = -n_p e \mathbf{E} - n_p m_0 \mathbf{g} + n_p \mathbf{\nabla} \mu_p.
	\label{eq:d_p}
\end{align}

After this point, the full derivation for the four-component mixture becomes rather lengthy.
However, following Ref.~\cite{GC_2020} we impose the condition of diffusive equilibrium for unbound neutrons: $\mathbf{\nabla} \mu_n = m_0 g$ and, by analogy, the same for unbound protons: $\mathbf{\nabla} \mu_p = m_0 g + e \mathbf{E}$. 
Consequently, we have $\mathbf{d}_n = \mathbf{d}_p = 0$, and then, taking into account Eq.~\eqref{eq:e_equil}, one gets:  $\mathbf{d}_1 + \mathbf{d}_2 = 0$.
From this we can express the electric field
\begin{align}
	\left(Z_1 n_1 + Z_2 n_2 \right) e \mathbf{E} = n_1 \mathbf{\nabla} \mu_1 + n_2 \mathbf{\nabla} \mu_2 - m_0 \mathbf{g} \left(A_1 n_1 + A_2 n_2 \right).
	\label{eq:E} 
\end{align}
Substituting Eq.~\eqref{eq:E} into Eq.~\eqref{eq:d_1} after some math, one obtains: 
\begin{align}
	\mathbf{d}_1 = \frac{n_1 n_2}{Z_1 n_1 + Z_2 n_2} \left[\left( \frac{A_2}{Z_2} - \frac{A_1}{Z_1} \right) Z_1 Z_2 m_0 g + Z_2\mathbf{\nabla} \mu_1 - Z_1 \mathbf{\nabla} \mu_2 \right],
	\label{eq:d_1f}
\end{align}
i.e., exactly the same as Eq.~(11) in Ref.~\cite{BY_2013}.
Eqs.~\eqref{eq:J_1} and \eqref{eq:J_2} reduce to Eq.~(8) of Ref.~\cite{BY_2013} as well. 
So, the question of the stability of the inner crust with respect to diffusion of ions (buoyancy) is still valid despite the presence of unbound nucleons.
Eqs.~(12) and (18) of Ref.~\cite{BY_2013} remain valid for the inner crust. 
Eq.~(16) of Ref.~\cite{BY_2013} is valid at least to some extent~\footnote{In our eNSE model, the chemical potentials are  more complicated than just ideal and Coulomb terms due to the excluded volume approximation that couples different subsystems.} as long as the plasma is strongly coupled, which is usually the case in the inner crust.
However, Eq.~(19) of Ref.~\cite{BY_2013} is no longer valid as it relies on the fact that in the outer crust the pressure is dominated by the pressure of electrons, which is no longer the case for the inner crust. 

So, from Eq.~\eqref{eq:d_1f} one can see that the behavior of the $\slfrac{A}{Z}$ ratio with density determines the sign of the diffusive flux. 
Of course, the second term, i.e., $Z_2\mathbf{\nabla} \mu_1 - Z_1 \mathbf{\nabla} \mu_2$, can counteract the first term.    
Our estimates demonstrated, however, that, for the densities close to the transition density, the first term dominates the flux. 
Moreover, not only $\slfrac{A}{Z}$ may decrease with density in some models, but $Z$ as well. In that case, both first and second terms give the contribution of the same sign.  

\bibliography{Crust.bib}
\end{document}